\documentclass[journal]{IEEEtran}

\usepackage{epsfig}

\usepackage{color}
\usepackage{latexsym,amssymb,amsmath,setspace,array,multirow}
\usepackage{algorithmic}
\usepackage{algorithm}
\usepackage{graphicx}
\usepackage{booktabs}
\usepackage{cases}
\usepackage{color}
\usepackage{subfig}
\usepackage{caption}

\usepackage{makecell}
\usepackage{mathrsfs}
\usepackage{amsbsy}
\usepackage{comment}
\usepackage{ulem}
\usepackage[figuresright]{rotating}
\newcommand{\degree}{^\circ}

\begin{document}
%	\title{Deep Neural Networks with Koopman Operators for Modeling and Control of Autonomous Vehicles}
	
\title{Deep Neural Networks with Koopman Operators for Modeling and Control of Autonomous Vehicles}
\author{Yongqian~Xiao,
	Xinglong~Zhang, \IEEEmembership{Member, IEEE},
	Xin~Xu, \IEEEmembership{Senior Member, IEEE}, 
	Xueqing~Liu, and
	Jiahang Liu
		\thanks{This work has been submitted to the IEEE for possible publication. Copyright may be transferred without notice, after which this version may no longer be accessible.}
		\thanks{Yongqian Xiao, Xinglong Zhang, Xin Xu, Xueqing Liu, and Jiahang Liu are with the College of Intelligence Science and Technology, National University of Defense Technology, Changsha 410073, China. email: (xiaoyongqian18@nudt.edu.cn; zhangxinglong18@nudt.edu.cn; xuxin\_mail@263.net; 13795351841@163.com; liujiahang1992@foxmail.com)}% <-this % stops a space
}
	
%	\markboth{Journal of \LaTeX\ Class Files}%,~Vol.~14, No.~8, August~2015}%
%	{Shell \MakeLowercase{\textit{et al.}}: Deep Neural Networks with Koopman Operator for Data-driven Modeling of Vehicle Dynamics}
	



	% for Transactions on Magnetics papers, we must declare the abstract and
	% index terms PRIOR to the title within the \IEEEtitleabstractindextext
	% IEEEtran command as these need to go into the title area created by
\maketitle
	\begin{abstract}
			Autonomous driving technologies have received notable attention in the past decades. In autonomous driving systems, identifying a precise dynamical model for motion control is nontrivial due to the strong nonlinearity and uncertainty in vehicle dynamics. Recent efforts have resorted to machine learning techniques for building vehicle dynamical models, but the generalization ability and interpretability of existing methods still need to be improved. In this paper, we propose a data-driven vehicle modeling approach based on deep neural networks with an interpretable Koopman operator. The main advantage of using the Koopman operator is to represent the nonlinear dynamics in a linear lifted feature space. In the proposed approach, a deep learning-based extended dynamic mode decomposition algorithm is presented to learn a finite-dimensional approximation of the Koopman operator. Furthermore, a data-driven model predictive controller with the learned Koopman model is designed for path tracking control of autonomous vehicles. Simulation results in a high-fidelity CarSim environment show that our approach exhibit a high modeling precision at a wide operating range and outperforms previously developed methods in terms of modeling performance. Path tracking tests of the autonomous vehicle are also performed in the CarSim environment and the results show the effectiveness of the proposed approach.
		\end{abstract}
		

%	\maketitle

%	\IEEEdisplaynontitleabstractindextext

	% \fi
%	\IEEEpeerreviewmaketitle
	
	\section{Introduction}   \label{doa:sec:introduction}
	%******************************************************************
	\IEEEPARstart{A}{utonomous} vehicles and driving technologies have received notable attention in the past several decades. Autonomous vehicles are promising to free the hands of humans from tedious long-distance drivings and have huge potential to reduce traffic congestion and accidents. A classic autonomous driving system usually consists of key modules of perception, localization, decision-making, trajectory planning and control. In trajectory planning and control, \textcolor{black}{the} information of vehicle dynamics is usually required for realizing agile and safe maneuver, especially in complex and unstructured road scenarios.
	
	In order to realize high-performance trajectory tracking of autonomous vehicles, many control methods, such as Linear Quadratic Regulator (LQR)~\cite{chen2019autonomous} and Model Predictive Control (MPC)~\cite{moser2017flexible}, rely on vehicle model information with different levels and structures. However, to obtain a precise model of vehicle dynamics is difficult for the following reasons: i) the vehicle dynamics in the longitudinal and lateral directions are highly coupled; ii) strong non-linearity and model uncertainty become dominant factors especially when the vehicle reaches the limit of tire-road friction \cite{rajamani2011vehicle, pacejka2005tire, Dieter2014vehicledynamics}. Among the existing approaches, classic physics-based modeling methods rely on Newton's second law of motion, where multiple physical parameters are required to be determined, see~\cite{Vicente2020,Berntorp2019}. Indeed, some model parameters, such as cornering stiffness coefficient, are not measurable and difficult to be estimated \cite{sierra2006cornering, vicente2020linear,na2017active}.

As an alternative to analytic modeling of vehicle dynamics, machine learning techniques have been used in recent years for building data-driven models of vehicles, see~\cite{spielberg2019neural,deo2018convolutional}. However, deep neural networks commonly lack interpretability, which has recently been noted as a challenge for applications with safety requirements and remains as a cutting-edge research topic. As a result, the models established by deep neural networks might have unknown sensitive modes and be fragile to model uncertainties. Also, due to the nonlinear activation functions, the obtained dynamics model is not easy to be used for designing a well-posed controller such as MPC, LQR, and so forth. Recently, the Koopman operator, being an invariant linear operator (probably of infinite dimension), has been regarded as a powerful tool for capturing the intrinsic characteristics of a nonlinear system via linear evolution in the lifted observable space. To obtain a realistic model description, usually, dimensionality reduction methods such as Dynamic Mode Decomposition (DMD) in \cite{schmid2010dynamic,kevrekidis2016kernel} and Extended Dynamic Mode Decomposition (EDMD) in \cite{williams2015data} can be used to approximate the Koopman operator with a finite dimension matrix.  In~\cite{cibulka2019data},  a data-driven identification method using Koopman operator was proposed for vehicle dynamics but the basis functions were either manually designed or determined by the knowledge of the nonlinear dynamics, which is a nontrivial task due to the dynamics in the longitudinal and lateral directions are strongly coupled and highly nonlinear. It is necessary to study learning-based construction methods for basis functions of the Koopman operator. Furthermore, the data-driven vehicle models studied in previous works have not been verified in practical controller design process. %for and no control validation is performed for the learned dynamics.} %\textcolor{red}{Previous methods based on the Koopman operator either required to design the basis functions manually, resulting in poor accuracy, or did not apply to forced systems, or required to calculate the system matrix by least-squares corresponding to the batch of input data, which made neural networks difficult to converge when the systems are complex. Besides, neural network-based methods struggle with interpretability and resulting in a non-linear model that brings inconvenience for motion planning and control of vehicles with optimization approaches, such as MPC.}
	
	To solve the above problems, this paper proposes a novel data-driven vehicle modeling approach based on deep neural networks with an interpretable Koopman operator. In the proposed approach, a deep learning-based extended dynamic mode decomposition (Deep EDMD) algorithm is presented to learn a finite basis function set of the Koopman operator. Furthermore, a data-driven model predictive controller is designed for path tracking control of autonomous vehicles by making use of the dynamic model learned by Deep EDMD. In the proposed algorithm, deep neural networks serve as the encoder and decoder in the framework of EDMD for learning the Koopman operator. The simulation studies in a high-fidelity CarSim environment are performed for performance validation. The results show that the proposed Deep EDMD method outperforms EDMD, deep neural network (DNN), and extreme learning machine-based EDMD (ELM-EDMD, designed using the ELM described in \cite{huang2006extreme}) algorithms in terms of modeling performance.  Also, the model predictive controller designed with Deep EDMD can realize satisfactory trajectory tracking performance and meet real-time implementation requirements.
	
	The contributions of the paper are two-fold:
	
	1) An EDMD-based deep learning approach (Deep EDMD) is proposed for identifying an integrated vehicle dynamical model in both longitudinal and lateral directions.
	Different from other machine learning-based approaches, deep neural networks play the role of learning feature representations for EDMD in the framework of the Koopman operator. Also, the deep learning network is utilized for automatically learning suitable observable basis functions, which is in contrast to that being manually selected or based on prior model information as in~\cite{cibulka2019data}. The resulting model is a linear global model in the lifted space with a nonlinear mapping from the original state space.
	
	2) A novel model predictive controller was designed with the learned model using Deep EDMD (called DE-MPC) for time-varying reference tracking of autonomous vehicles. In DE-MPC, the linear part of the model is used as the predictor in the prediction horizon of the MPC, while the nonlinear mapping function is used for resetting the initial lifted state at each time instant with the real state. 
		
	The rest of this paper is structured as follows. In section \uppercase\expandafter{\romannumeral2}, the most related works are reviewed, while the dynamics of four-wheel vehicles are shown in Section \uppercase\expandafter{\romannumeral3}. Section \uppercase\expandafter{\romannumeral4} presents the main idea of the Deep EDMD method for data-driven model learning. And Section \uppercase\expandafter{\romannumeral5} gives a linear MPC with Deep EDMD  for reference tracking. The simulation results are reported in Section\uppercase\expandafter{\romannumeral6}, while  some conclusions are drawn in Section \uppercase\expandafter{\romannumeral7}.
	
	\section{Related Works}
	In the past decades, neural networks were widely studied to approximate dynamical systems due to their powerful representation capability. Earlier works in \cite{funahashi1993approximation, li2002dynamic} have resorted to recurrent neural networks (RNNs) for the approximation of dynamics. It was verified that any continuous curve can be represented by the output of RNNs. In~\cite{garimella2017neural}, an RNN-based modeling approach was presented to model the dynamical steering behavior of an autonomous vehicle. The obtained nonlinear model was used to design a nonlinear MPC for realizing steering control. In addition to RNNs, fruitful contributions have resorted to artificial neural networks (ANNs) with multiple hidden layers for the approximation of vehicle dynamics. Among them, \cite{spielberg2019neural} used fully-connected layers to model the vehicle dynamics in the lateral direction. The physics model was considered for determining the inputs of the networks to guarantee the reliability of the modeling results. It is noteworthy that most of the models with neural networks are nonlinear, and the identification process may suffer from the vanishing of gradients. In recent years, the reliability in identification with neural networks has received notable attention. In \cite{da2019modelling}, different levels of convolution-like neural network structures were proposed for modeling the longitudinal velocity of vehicles. During the construction of neural networks, it integrated the physics principles to improve the robustness and interpretability of the modeling. Note that, most of the above approaches assumed the decoupling of dynamics and approximate in the sole longitudinal direction or lateral direction. As a fact, couplings between both directions can be naturally dealt with when it comes to use multi-variable control techniques, such as MPC and LQR.
	
	As an alternative, the Koopman operator has been recently realized to be a powerful tool for representing nonlinear dynamics. The main idea of the Koopman operator was initially introduced by Koopman~\cite{koopman1931hamiltonian}, with the goal of describing the nonlinear dynamics using a linear model with the state space constructed by observable functions. It has been proven that the linear model with the Koopman operator can ideally capture all the characteristics of the nonlinear system as long as the state space adopted is invariant. That is to say, the structure with the Koopman operator is interpretable. A sufficient condition for the convergence of the Koopman operator is that an infinite number of observable functions are provided.

For practical reasons, dimensionality reduction methods, such as DMDs, have been used to obtain a linear model of finite order, see \cite{schmid2010dynamic,tu2013dynamic} and the references therein. The order of the resultant model corresponds to the number of observable functions used. In general, the observable functions are directly measured with multiple sensors. To avoid redundant sensors, efforts have been contributed to extending the idea of DMDs for constructing observable functions, see~\cite{korda2018linear, li2017extended,williams2015data}. In EDMDs, the observable functions are designed using basis functions probably of multiple types, such as Gaussian functions, polynomial functions, and so forth. The convergence has been proven in~\cite{korda2018convergence} under the assumption that the state space constructed by basis functions is invariant or of infinite dimension.

	From a practical viewpoint, the approximation performance might be sensitive to basis functions, hence the selection of basis functions requires specialist experience. Motivated by the above problems, recent works have resorted to design deep neural networks to generate observable functions automatically. In \cite{otto2019linearly}, an autoencoder is utilized to learn dictionary functions and Koopman modes for unforced dynamics. The finite-horizon approximation of the Koopman operator was computed via solving a least-squares problem. In \cite{morton_deep_2019}, a deep variational Koopman (DVK) model was proposed with a Long Short Term Memory (LSTM) network for learning the distribution of the lifted functions. A deep learning network was used in \cite{lusch2018deep} to learn Koopman modes in the framework of DMD. From the application aspect, a deep DMD method has been proposed for the modal analysis of fluid flows in~\cite{morton2018deep}. In~\cite{susuki2016applied} and~\cite{susuki2013nonlinear}, the Koopman operator has been extended to the modeling of power systems. In~\cite{avila2020data}, a highway traffic prediction model was built based on the Koopman operator. Other applications can be founded for representing neurodynamics in \cite{brunton2016extracting} and molecular dynamics in \cite{wu2017variational, mardt2018vampnets}, and model reduction in \cite{peitz2019koopman}.
	In the aspect of robot modeling and control,  a DMD-based modeling and controlling approach was proposed for soft robotics in~\cite{bruder2019modeling} and the Koopman operator with EDMD was applied to approximate vehicle dynamics in~\cite{cibulka2019data}.  Different from the above works, we propose a deep neural network to learn the basis functions of the Koopman operator in the framework of EDMD, with the goal of data-driven modeling of vehicles in both longitudinal and lateral directions. The deep neural networks serve as the encoder and decoder whose weights are automatically updated in the learning process, avoiding the manual selection problem of observable functions as in~\cite{cibulka2019data}. %To the best of our knowledge, no prior contribution has been presented \sout{in this respect for approximating vehicle dynamics}\textcolor{red}{for approximating vehicle dynamics with deep neural networks based on EDMD}.
	
	\section{The Basic Model Description of Autonomous Vehicles}
\begin{figure}[htbp]
		\begin{center}
			\includegraphics[scale=0.45]{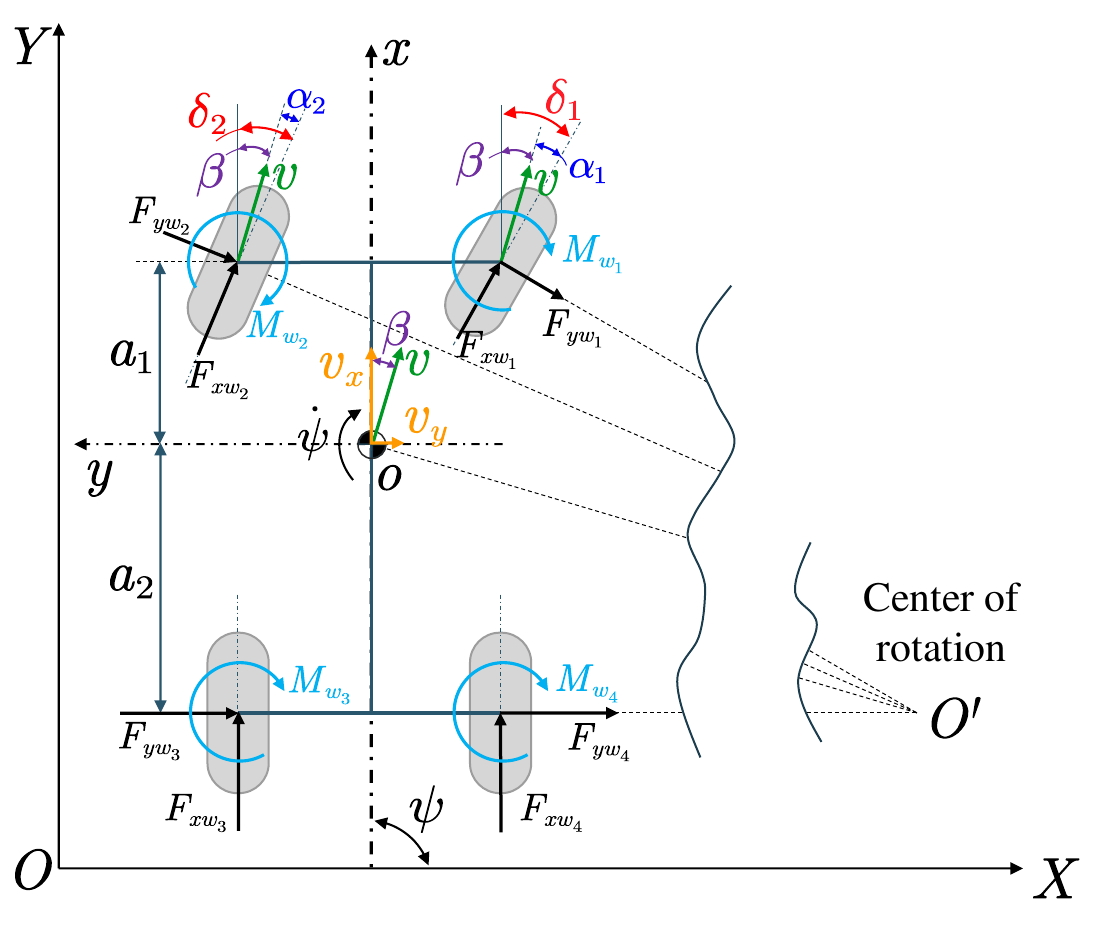}
			\caption{A planar four-wheel model of a front-wheel-steering vehicle with no roll motion and the forces and moments acting on the tires.
			}
			\label{fig:vehicle_dynamics}
		\end{center}
	\end{figure}

	In this paper, we consider the type of autonomous vehicles consisting of four wheels with front-wheel-steering functionality. A sketch of the vehicle dynamics is depicted in Figure~\ref{fig:vehicle_dynamics}.
	
	As shown in Figure \ref{fig:vehicle_dynamics}, according to~\cite{rajamani2011vehicle,jazar2017vehicle}, the longitudinal and lateral forces can be represented as
	\begin{equation}\label{equ:longitudinal_lateral_forces}
	%\begin{array}
	\begin{array}{l}
	\left[ \begin{array}{l}
	F_x\\
	F_y\\
	\end{array} \right] = \sum_{i=1}^4{\left[ \begin{array}{c}
		F_{x_i}\\
		F_{y_i}\\
		\end{array} \right]}\\
	=\sum_{i=1}^4{\left[ \begin{matrix}
		\cos \delta _i&		-\sin \delta _i\\
		\sin \delta _i&		\cos \delta _i\\
		\end{matrix} \right]}\left[ \begin{array}{c}
	F_{xw_i}\\
	F_{yw_i}\\
	\end{array} \right]-\frac{1}{2}\rho AV^2\left[ \begin{array}{c}
	c_{fx}\\
	c_{fy}\\
	\end{array} \right]\\
	\end{array}
	\end{equation}
	where $F_{xw_i}$, $F_{yw_i}$ denote the longitudinal and lateral forces of tire $i$, $\delta _i$ denotes the wheel angle of tire $i$, and $\delta_3=\delta_4=0$. The term $-\frac{1}{2}\rho AV^2\left[ \begin{array}{c}
	c_{fx}\\
	c_{fy}\\
	\end{array} \right]$ denotes the longitudinal and lateral air-resistance forces applied to the vehicle body. The parameter $\rho$, $ A$, and $ V$ denote the air mass density, the frontal area to the airflow, and the relative air speed respectively, and where $c_{fx}$ and $c_{fy}$ are drag coefficients defined as configurable functions with respect to the yaw angle. $F_x$ and $F_y$ can be calculated by the Pacejka model in \cite{pacejka2005tire,jazar2017vehicle}, which is an empirical formula given as
	\begin{equation}\label{pacejka_model}
	\begin{aligned}
	F_x=H\sin \left\{ E\tan ^{-1}\left[ Gs-D\left( Gs-\tan ^{-1}\left( Gs \right) \right) \right] \right\}
	\end{aligned}
	\end{equation}
	where $H=\varLambda_{\ast} F_z$, and where $F_z$ is the tire load while $\varLambda$ is a function with respect to the longitudinal slip $s$ and sideslip angle $\alpha$. Besides, $s$ and $\alpha$ are functions with respect to the steering wheel angle $\zeta$ and the engine $\eta$ (stands for the throttle if $\eta\geq0$ and for the brake otherwise). That is to say, $F_x$ is a function with respect to $\zeta$ and $\eta$, i.e. $s=\vartheta \left(\zeta, \eta \right)$. $E$ and $D$ are shape factors and $G=\frac{C_{\alpha}}{HE}$, where $C_{\alpha}$ is the cornering stiffness. According to the Newton-Euler equation, we can obtain equations of motion as follows:
	\begin{subequations}
		\begin{align}\label{equ:v_x_dot}
		\dot{v}_x=\frac{1}{m}F_x+\dot{\psi}v_y
		\end{align}
		\begin{align}\label{equ:v_y_dot}
		\begin{array}{ll}
		\dot{v}_y=& \frac{1}{mv_x}\left( -a_1C_{\alpha f}+a_2C_{\alpha r} \right) \dot{\psi} +\frac{1}{m}C_{\alpha f}\delta\\
		& -\dot{\psi}v_x -\frac{1}{mv_x}\left( C_{\alpha f}+C_{\alpha r} \right) v_y
		\end{array}
		\end{align}
		\begin{align}\label{equ:yaw_rate}
		\begin{array}{ll}
		\dot{\psi}= & \frac{1}{I_zv_x}\left( -a_{1}^{2}C_{\alpha f}-a_{2}^{2}C_{\alpha r} \right) \psi +\frac{1}{I_z}a_1C_{\alpha f}\delta \\ &-\frac{1}{I_zv_x}\left( a_1C_{\alpha f}-a_2C_{\alpha r} \right) v_y
		\end{array}
		\end{align}
	\end{subequations}
	where $v_x$ and $v_y$ are the longitudinal and lateral velocities respectively, $m$ is the mass of vehicle, $I_z$ is the yaw moment of inertia, $\psi$ is the yaw angle, $\beta$ is the slip angle, $\delta = \frac{\delta_1 + \delta_2}{2}$, $a_1$ and $a_2$ are the distances between the front and rear axles to the center of gravity, $C_{\alpha f}=C_{\alpha f L} + C_{\alpha f R}$ and $C_{\alpha r}=C_{\alpha r L} + C_{\alpha r R}$.
	
	Combining~\eqref{equ:longitudinal_lateral_forces}-\eqref{equ:yaw_rate}, it is possible to write the vehicle dynamics with interactions in the longitudinal and lateral directions, that is
	\begin{equation}\label{equ:vehicle_dynamics}
	\dot x=f_c(x,u)
	\end{equation}
	where the state $x=(v_x,v_y,\dot{\psi})$, $u=(\zeta,\eta)$, where the steering wheel angle $\zeta$ establishes a relation with $\delta_i$, i.e., $\delta=\varUpsilon(\zeta)$.  The choice of dynamics constructed as~\eqref{equ:vehicle_dynamics} avoids to measure $F_x$, $F_y$, and $F_z$ with expensive sensors, but it is challenging from the modeling viewpoint due to the strong non-linearity being hidden in $f_c$. For this reason, we propose a deep neural network method based on the Koopman operator to capture the global dynamics of~\eqref{equ:vehicle_dynamics} with a linear model using lifted abstract state space.
	
	\section{The Deep EDMD Approach for Vehicle Dynamics Modeling}
	In this section, we first give the formulation with the Koopman operator and its numerical approximation with EDMD for approximating vehicle dynamics. Then, we present the main idea of the Deep EDMD algorithm for modeling the vehicle dynamics, where a deep neural network is utilized to construct an observable subspace of the Koopman operator automatically.
	
	\subsection{The Koopman operator using EDMD}
	The Koopman operator has been initially proposed for capturing the intrinsic characteristics via a linear dynamical evolution for an unforced nonlinear dynamics. With a slight change, the Koopman operator can also be used for representing systems with control forces. The rigorous definition of the Koopman operator can be defined as in~\cite{schmid2010dynamic,tu2013dynamic}. Instead, we formulate our modeling problem to fit with the Koopman operator. Let
	\begin{equation}\label{Eqn:mo-de}
	x_{k+1}=f(x_k,u_k)
	\end{equation}
	be the discrete-time version of~\eqref{equ:vehicle_dynamics} with a specified sampling interval, where $k$ is the index in discrete-time.
	Let $z=(x,\boldsymbol{u})$, where $\boldsymbol{u}=u_0^{\infty}$, and $\boldsymbol{u}=\Gamma \boldsymbol{u}$, $\Gamma$ is a shift operator. One can define the Koopman operator on~\eqref{Eqn:mo-de} with the extended complete state space being described as% an observable $\varphi(x)\in \mathcal{O}$, that is
	\begin{equation}\label{equ:base_Koopman}
	(\mathcal{K}\varphi){(z_k)}=\varphi\ \circ \ f = \varphi(z_{k+1})
	\end{equation}
	where $\mathcal{K}$ denotes the Koopman operator, $\circ$ is the composition of $\varphi$ with $f$, $\varphi(z)\in \mathcal{O}$ is the observable in the lifted space. \eqref{equ:base_Koopman} can be seen as representing dynamics~\eqref{equ:vehicle_dynamics} via the Koopman operator with a lifted (infinite) observable space $\mathcal{O}$ via a linear evolution.
As the state $z$ is of infinite horizon, for practical reasons we adopt $z=(x,u)$  in the rest of this paper for approximating the Koopman operator with a finite horizon. In the classic Koopman operator, the observables $\varphi$ are eigenfunctions associated with eigenvalues $\mu$, which reflects the intrinsic characteristics of the nonlinear dynamics.
Let  $\mathcal{K}$ be \textcolor{black}{approximated with} $N$ Koopman eigenfunctions $\varphi_1,\cdots,\varphi_N$ and modes $\xi_1, \cdots, \xi_N$ corresponding to the eigenvalues $\mu_1,\cdots,\mu_N$. One can predict the nonlinear system at any time $k$ in the lifted observable space as follows~\cite{otto2019linearly}:
	\begin{equation}\label{equ:classic_Koopman}
	\begin{aligned}
	x(k)= \sum_{i=1}^N{{\xi}_i(\mathcal{K}^k {\varphi}_i)({x_0, u_0})}  = \sum_{i=1}^N {{\xi }}_i \mu _{i}^k {\varphi} _i \left( {x_0, u_0} \right)                                                                                  \\
	\end{aligned}
	\end{equation}
\\   The main idea of EDMD for approximating the Koopman operator is to represent the observable functions with basis functions and compute a finite-horizon version of the Koopman operator using the least-squares method. In this case, the observable functions can be designed with multiple types of weighted basis functions, e.g. radial basis functions (RBF) with different kernel centers and widths:
	\begin{equation}\label{equ:Koopman-basis}
	\varphi(z) =[\phi(x)^{\top}\ u^{\top}] \boldsymbol{a}
	\end{equation}
where ${\phi(x)}=[\phi_1({x})\ \phi_2({x})\ ...\ \phi_L({x})]^{\top}$, $\phi_i$, $i=1,\cdots,L$,   are basis functions, and where $L=N-m$, and $\boldsymbol{a} \in \mathbb{R}^{N \times N}$ is the weighting matrix. %$\varphi(x) \in \mathbb{R}^{K \times 1}$ can ba regarded as the Koopman eigenfunctions. %Therefore, EDMD has better approximating capacity in theory while we choose proper lifted functions.
	With~\eqref{equ:Koopman-basis},~\eqref{equ:base_Koopman} becomes
	\begin{equation}\label{equ:Koopman-K}
	\mathcal{K} \varphi(z)=(\varphi^{\top} \circ f) \boldsymbol{a}=[\phi(x)^{\top}\ u^{\top}] {K} \boldsymbol{a}+\boldsymbol{r}
	\end{equation}
where $\boldsymbol{r}$ is a residual term. To optimize $K \in \mathbb{R}^{N \times N}$, $\|\boldsymbol{r}\|_F^2$ is considered as the cost to be minimized. For forced dynamics, i.e. the vehicle dynamics (\ref{equ:vehicle_dynamics}), we define $[A\ B]$ where $A\in\mathbb{R}^{L\times L}$ and $B\in\mathbb{R}^{L\times m}$ corresponding to the former $L$ lines of $K$ and $C\in \mathbb{R}^{m \times L}$ be the former $L$ columns of the last m lines of $K$. Therefore, the lifted vehicle dynamics based on the Koopman operator can be written as
	\begin{equation}\label{equ:classic_EDMD}
	\left\{\begin{array}{ll}
	\phi(x_{k+1})=[A\ B]\varPsi(x_{k})\\
	\hat{x}_{k}=C \phi(x_{k})
	\end{array}\right.
	\end{equation}
where $\varPsi(x)=[\phi(x)^{\top}\ u^{\top}]^{\top} \in \mathbb{R}^{K \times 1}$. The analytical solution of $A$, $B$ can be computed via minimizing the residual term $\boldsymbol{r}$
	\begin{equation}\label{equ:loss_EDMD}
	\begin{aligned}
	{\min_{A,B}}J ={\min_{A,B}}\lVert \boldsymbol{r} \rVert _{F}^{2} =\mathop {\min} \limits_{A,B}\sum_{k=1}^M{\lVert \phi \left( y_k \right) -\left[ A\,\,B \right] \varPsi \left( x_k \right) \rVert _{F}^{2}}
	\end{aligned}
	\end{equation}
where $y_k=f(x_k, u_k)$. Its analytical solution is
	\begin{equation}\label{equ:solve_1}
	\left[ A\,\,B \right] =VW^{\top}\left( WW^{\top} \right) ^{\dagger}
	\end{equation}
where $W=\left[ \varPsi \left( x_1 \right) \,\,\varPsi \left( x_2 \right) \cdots \,\,\varPsi \left( x_M \right) \right] $, $V=\left[ \phi \left( y_1 \right) \,\,\phi \left( y_2 \right) \cdots \,\,\phi \left( y_M \right) \right] $. Similarly, we can obtain the matrix $C$ by solving:
	\begin{equation}\label{equ:loss_Modes}
	\mathop {\min} \limits_{C}\sum_{k=1}^M{\lVert x_k-C\phi \left( x_k \right) \rVert _{F}^{2}}
	\end{equation}
	leading to
	\begin{equation}\label{equ:find_C}
	C=X\varXi ^{\top}\left( \varXi \varXi ^{\top} \right) ^{\dagger}
	\end{equation}
where $X=\left[ x_1\,x_2\,\,\cdots \,x_M \right]$, $\varXi =\left[ \phi \left( x_1 \right) \,\phi \left( x_2 \right) \,\cdots \,\phi \left( x_M \right) \right]$.
	
	\subsection{The Deep EDMD Algorithm}
	Different from EDMD, in this subsection we propose a Deep EDMD algorithm for modeling the vehicle dynamics, in which a deep neural network is utilized to construct an observable subspace of the Koopman operator automatically. Also, the merit of Deep EDMD with respect to classic multi-layer perceptions is that, the deep learning network is integrated into the EDMD method to learn an estimate of the Koopman operator with finite dimension. To proceed, one can write the approximated dynamics resulting from Deep EDMD in the following form:
	\begin{equation}\label{equ:Deep_EDMD}
	\left\{\begin{array}{ll}
	\boldsymbol\phi_{e,t+1}&=K\varPsi(x_t,\theta_{e},u)\\
	\hat{x}_{t}&=\tilde{\varPsi}(\boldsymbol\phi_{e,t},\theta_{d})
	\end{array}\right.
	\end{equation}
	where $K= [\mathscr{A} \ \mathscr{B}] \in \mathbb{R}^{L \times N}$, $\varPsi=[\boldsymbol\phi_e (x,\theta_{e})^{\top}\ u^{\top}]^{\top} \in \mathbb{R}^{N}$, $\boldsymbol\phi_e$ is the encoder parameterized with $\theta_{e}$, $\tilde{\varPsi}(\cdot)$ is the decoder parameterized with $\theta_d$. With a slight abuse of notations, in the rest of the paper we use $\varPsi_t$  to stand for $\varPsi(x_t,\theta_{e},u)$ unless otherwise specified.
	
	\begin{figure*}[htbp]
		\begin{center}
			\includegraphics[scale=0.6]{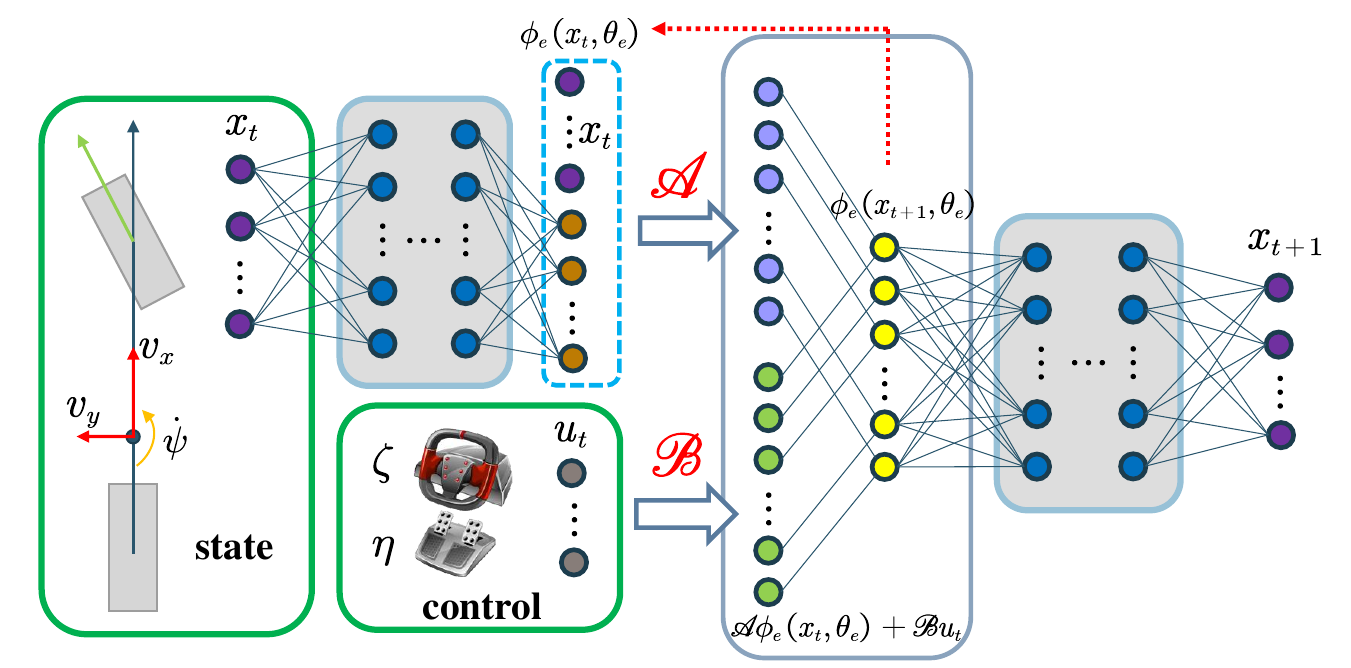}
			\caption{The diagram of the proposed Deep EDMD. The original state $x_t$ is lifted with the encoder, i.e., $\phi_e(x_t,\theta_e)$. Then $\phi_e(x_t,\theta_e),\ u_t$ forms the lifted state for constructing the linear evolution in the lifted space. The vehicle dynamics can be recovered via a decoder from the lifted state space.  %Deep EDMD bases on a deep auto-encoder. The encoder lifts the original vehicle state $x_t$ to a high-dimensional state to adaptively learn the basic functions and the decoder maps it back to the original state space for taking the role of Koopman modes. Especially, we take the state and input matrices $\mathscr{A}$ and $\mathscr{B}$ as fully-connected (FC) layers of the network. Therefore, we can obtain the $\mathscr{A}$ and $\mathscr{B}$ according to the corresponding network weights after training. Note that the layers of $\mathscr{A}$ and $\mathscr{B}$ without any activation so that we can make sure the system is linear on the lifted state space $\phi$.
			}
			\label{fig:DeepEDMD_network}
		\end{center}
	\end{figure*}
	As shown in Fig. \ref{fig:DeepEDMD_network}, the Deep EDMD algorithm relies on an autoencoder structure. The encoder, i.e., $\phi_e$, consisting of fully-connected layers, is in charge of mapping the original state to a lifted observable space. The weights $\mathscr{A}$ and $\mathscr{B}$ are placed connecting to the last layer of the encoder without activation functions. In principle, $\mathscr{A}$ and $\mathscr{B}$ can be trained synchronously with the encoder. In case the encoder might experience unanticipated errors and vanishing gradient, $\mathscr{A}$ and $\mathscr{B}$ can be updated as the last training step in which way, the worst scenario of Deep EDMD might degrade into EDMD. {Similar to the encoder, the decoder consists of fully-connected layers,} devoting to recovering the original state from the lifted observable space. To be specific, at any hidden layer $l$, the output can be described as
	\begin{equation}\label{Eqn:layer-eq}
	\mathscr{Y}_{\ast}^{\left( l \right)}=\sigma_{\ast}^{\left( l \right)}\left( W_{\ast}^{\left( l \right)} \mathscr{Y}_{\ast}^{\left( l-1 \right)}+b_{\ast}^{\left( l \right)} \right)
	\end{equation}
where $*=e,d$ in turn stands for the subscripts for encoder or decoder, $W^{(l)} \in \mathbb{R}^{n_{l}\times n_{l-1}}$ and $b^{(l)} \in \mathbb{R}^{n_l}$ are the weight and bias of the hidden layer $l$, where $\sigma^{(l)}$ denotes the activation functions of the hidden layer $l$ of the encoder and decoder, $l=1,\cdots, H$, where $H$ is the number of layers of the encoder and decoder. In the encoder, $\sigma_e^{l}$, $l=1,\cdots,H-1$, is designed using a rectified linear units (ReLU), see \cite{nair_rectified_2010}, while no activation function is used in the last layer. As for the decoder, $\sigma_d^{l}$, $l=1,\cdots,H-1$ uses ReLU as the activation, while the last layer adopts a Sigmoid activation function. In this way, the lifted state and the predicted state can be computed with the last layer of the parameterized networks, i.e.,
	\begin{equation}
	\phi_e \left( x, \theta_{e} \right) =W_e^{\left( H \right)}\mathscr{Y}_e^{\left( H-1 \right)}
	\label{equ:encoder_parameterized}
	\end{equation}
	\begin{equation}
	\hat{x}= \tilde{\varPsi}\left( \phi_e,\theta _{d} \right)  = \text{sig} \left( W^{\left( H \right)}\mathscr{Y}_d^{\left( H-1 \right)}+b^{\left(H \right)} \right)
	\label{equ:decoder_parameterized}
	\end{equation}
	
	Concerning~\eqref{Eqn:layer-eq}, one can promptly get the whole expressions of the encoder from~\eqref{equ:encoder_parameterized} as well as the decoder from~\eqref{equ:decoder_parameterized}.
	
	As the objective is to approximate the vehicle dynamics in a long time window, the multi-step prediction error instead of the one-step one is to be minimized. Hence the state and control sequences with time information are used to formulate the optimization problem. Different from that in EDMD with one-step prediction approximation, the resulting problem is difficult to be solved analytically, but it can be trained in a data-driven manner. To introduce the multi-step prediction loss function, we first write the state prediction in $p$ time steps:
	\begin{equation}
	x_{t+p} = \tilde{\varPsi}(K^{[p]}\varPsi_t,\theta_d) + \boldsymbol{r}_{x,p}
	\end{equation}
where $K^{[p]}\varPsi_t$ is the $p$-step ahead state starting from $x_t$, i.e.,
	\begin{equation}\label{equ:detail_Koopman}
	\begin{aligned}
	K^{[p]}\varPsi_t&=\phi_e \left( x_{t+p},\theta_e \right)\\
	&=\mathscr{A}\phi_e \left( x_{t+p-1},\theta_e \right) +\mathscr{B}u_{t+p-1}\\
	&=\mathscr{A}\left( \mathscr{A}\phi_e \left( x_{t+p-2},\theta_e \right) +\mathscr{B}u_{t+p-2} \right) +\mathscr{B}u_{t+p-1}\\
	&=\mathscr{A}^2\phi_e \left( x_{t+p-2},\theta_e \right) +\mathscr{A}\mathscr{B}u_{t+p-2}+\mathscr{B}u_{t+p-1}\\
	&=\mathscr{A}^p\phi_e \left( x_t,\theta_e \right) +\sum_{i=1}^p{\mathscr{A}^{i-1}\mathscr{B}u_{t+p-i}}\\
	\end{aligned}
	\end{equation}
	
	Specifically, we minimize the sum of prediction errors along the $p$ time steps, which leads to the loss function being defined as
	\begin{equation}
	\begin{aligned}
	\label{equ:loss_pred}
	L_{x,x}=\frac{1}{p}\sum_{i=1}^p{\lVert x_{t+i}-\tilde{\varPsi}\left( K^{[i]}\varPsi_t,\theta_d \right) \rVert _{2}^{2}}
	\end{aligned}
	\end{equation}
	
	Another perspective of validating the modeling capability of Deep EDMD is to evaluate the prediction error in the lifted observable space, i.e., it is crucial to minimize the error of the state evolution in the lifted space and the lifted state trajectory mapping from the real dynamics. To this objective, we adopt the loss function in the lifted linear space as
	\begin{equation}\label{equ:loss_line}
	L_{x,o}=\frac{1}{p} \sum_{i=1}^{p}{\lVert \phi_e (x_{t+i}, \theta_e) -K^{[i]}\varPsi_t \rVert}_{2}^{2}
	\end{equation}
	
	In order to minimize the reconstruction error, the following loss function about the decoder is to be minimized:
	\begin{equation}\label{equ:loss_recon}
	L_{o,x}=\frac{1}{p} \sum_{i=1}^{p}  \lVert x_i - \tilde{\varPsi} \left( \phi_e (x_i, \theta_e),\theta_d \right) \rVert _{2}^{2}
	\end{equation}
	
	Also, to guarantee the robustness of the proposed algorithm, the loss function in the infinite norm is adopted, i.e.,
	\begin{equation}\label{equ:loss_infinite}
	\begin{aligned}
	L_{\infty}=&\frac{1}{p}\sum_{i=1}^{p}{\lVert x_{i}-\tilde{\varPsi} (\phi_e (x_i, \theta_e),\theta_d) \rVert _{\infty}}+\\
	&\frac{1}{p}\sum_{i=1}^{p}{\lVert x_i-\tilde{\varPsi} \left( K^{[i]}\varPsi_0,\theta_d \right) \rVert _{\infty}}
	\end{aligned}
	\end{equation}
	
	With the above loss functions introduced, the resulting learning algorithm aims to solve the following optimization problem:
	\begin{equation}\label{Eqn:min-L}
	\min_{{\theta_e,\theta_d, \mathscr{A}, \mathscr{B}}}\  L
	\end{equation}
where $L$ is the overall optimization function defined as
	\begin{equation}\label{equ:loss_total}
	\begin{array}{ll}
	L = &\alpha_1 L_{o,x} + \alpha_2 L_{x,x} + \alpha_3 L_{x,o} +\alpha_4 L_{\infty} + \\
	
	&\alpha_5 \lVert {\theta}_e \rVert_{2}^{2}+\alpha_6 \lVert {\theta}_d \rVert_{2}^{2}
	\end{array}
	\end{equation}
where $\alpha_i$, $i=1,\cdots,6$, are the weighting scalars, $\lVert {\theta}_{\ast} \rVert_{2}^{2}$, $\ast=e,d$ in turn is a $l_2$ regularization term used for avoiding over-fitting.
	
	Assuming that the optimization problem~\eqref{Eqn:min-L} is solved, the resulting approximated dynamics for~\eqref{equ:vehicle_dynamics} can be given as
	\begin{equation}\label{equ:DeepEDMD_dynamics}
	\left\{ \begin{array}{l}
	\phi_e \left( x_{t+1} \right) =\mathscr{A} \phi_e \left( x_t,\theta _{e} \right) +\mathscr{B}u_t\\
	\hat{x}_t=\tilde{\varPsi} \left( \phi_e \left( x_t,\theta_e \right) ,\theta _{d} \right)\\
	\end{array} \right.
	\end{equation}
	
	\begin{algorithm}[h]
	\caption{The Deep EDMD Algorithm}
	\label{alg:Deep_EDMD}
	\text{// The training is performed in a batch-mode manner.}
	\begin{algorithmic}[1]
		\REQUIRE Initialize ${\theta}_e$, ${\theta}_d$, $\mathscr{A}$, $\mathscr{B}$, $p$, $Epoch = 0$, $Epoch_{max}$, $\alpha_i$, $i=1,\cdots,6$, batch size $b_s$, a small scalar $\epsilon>0$.\\
		\ENSURE trained ${\theta}_e$, ${\theta}_d$, $\mathscr{A}$, $\mathscr{B}$;\\
		\WHILE {$Epoch > Epoch_{max}$ or $|L|\leq \epsilon$}
		\STATE Reset the training episodes;
		\WHILE {$Epoch$ is not Terminated}
		\STATE Sample a batch data sequence of state and control, i.e., $X$, $U$;
		\STATE Obtain the lifted states $\phi_e(X,\theta_e)$ with (\ref{equ:encoder_parameterized}) and reconstruction states $\hat{X}=\tilde{\varPsi}(\phi_e(X,\theta_e),\theta_d)$ with (\ref{equ:decoder_parameterized});
		\STATE Compute multi-step lifted states $K^{[i]}\varPsi_0$ with (\ref{equ:detail_Koopman}) and predicted states $\tilde{\varPsi}(K^{[i]}\varPsi_0)$, where $i=1,2, \cdots, p$;
		\STATE Obtain the weighted loss $L$ with (\ref{equ:loss_total}), (\ref{equ:loss_recon}), (\ref{equ:loss_pred}), (\ref{equ:loss_line}), and (\ref{equ:loss_infinite});
		\STATE Update $\theta_e$, $\theta_d$, $\mathscr{A}$, and $\mathscr{B}$ via solving~\eqref{Eqn:min-L} with an Adam optimizer;
		\ENDWHILE
		\STATE $Epoch = Epoch + 1$
		\ENDWHILE
	\end{algorithmic}
\end{algorithm}
	
	\section{MPC with Deep EDMD for trajectory tracking of autonomous vehicles }
	In this section, a novel model predictive controller based on Deep EDMD, i.e., DE-MPC, is proposed to show how the learned dynamical model can be utilized for controlling the nonlinear system with good performance. To this end, the learned model~\eqref{equ:DeepEDMD_dynamics} is firstly used to define  an augmented version with the input treated as the extended state, i.e.,
	\begin{equation}\label{equ:model-mpc}
	\left\{ \begin{array}{l}
	\xi_{t+1} = \bar{\mathscr{A}}\xi_t +\bar{\mathscr{B}}\varDelta u_t\\
	{y}_t = \mathscr{C}\xi_t
	\end{array} \right.
	\end{equation}
	where $\Delta u_t=u_t-u_{t-1}$,
	\begin{equation}
	\begin{aligned}\nonumber
	\bar{\mathscr{A}}=\left[ \begin{matrix}
	\mathscr{A}&		\mathscr{B}\\
	0_{m\times L}&		I_{m\times m}\\
	\end{matrix} \right] ,\bar{\mathscr{B}}=\left[ \begin{array}{c}
	\mathscr{B}\\
	I_{m\times m}\\
	\end{array} \right]
	\end{aligned}
	\end{equation}
	$\mathscr{C}=\left[ \begin{matrix}
	I_{L\times L}&		0_{L\times m}\\
	\end{matrix} \right] ,\xi_t =\left[ \begin{array}{c}
	\chi_t\\
	u_{t-1}\\
	\end{array} \right]$.
	And the state initialization at any time instant is given as
	\begin{equation}\label{Eqn:initial}
	\chi_t= \phi_e ( x_t ,\theta_e)
	\end{equation}
	
	It is now ready to state a model predictive controller with~\eqref{equ:model-mpc} and~\eqref{Eqn:initial}. The objective of DE-MPC is to steer the time-varying reference $y_{ref}$, which can be computed with the lifted observable function using the reference trajectory.  To obtain satisfactory tracking performance, we also include the original state $x$ as the abstract one in the lifted space.
	Therefore, at any time instant $t$, the finite horizon optimization problem can be stated as
	\begin{subequations}
		\begin{align}\label{equ:MPC_ob_function}
		\begin{array}{ll}
		\underset{\varDelta u_t ,\varepsilon}{\min}J\left( \xi_t ,\varDelta u_t ,\varepsilon \right) =& \sum\limits_{i=1}^{N_p}{\lVert y_{t+i} -y_{ref,t+i} \rVert _{Q}^{2}}+\\
		&\sum\limits_{i=0}^{N_c-1}{\lVert \varDelta u_{t+i} \rVert_R^2 +\rho \varepsilon ^2}
		\end{array}
		\end{align}
		subject to:\\
		1) model constraint~\eqref{equ:model-mpc} with initialization~\eqref{Eqn:initial};\\
		2) the constraints on control and its increment:
		\begin{gather}
		u_{\min}<u_t <u_{\max}\\
		\varDelta u_{\min}-\varepsilon \boldsymbol{1}_{m}<\varDelta u_t <\varDelta u_{\max}+\varepsilon \boldsymbol{1}_{m}
		\end{gather}
	\end{subequations}
	\begin{comment}
	\begin{equation}\label{equ:aug_state_space_equation}
	\begin{aligned}
	z\left( k+1 \right) =&\tilde{\mathscr{A}}z\left( k \right) +\tilde{\mathscr{B}}\varDelta u\left( k \right) \\
	\tilde{y} =& \mathscr{C}z(k)
	\end{aligned}
	\end{equation}
	\end{comment}
	where $N_p$ is the prediction horizon, $N_c$ is the control horizon, $Q$, $R$ are positive-definite matrices for penalizing the tracking errors and the control increment; while {$\rho>0$ is the penalty matrix of the slack variable $\varepsilon$ and $\boldsymbol{1}_m\in\mathbb{R}^m$ is a vector with all the entries being $1$. Note that we use $N_p>N_c$ and the control increment $\Delta u_t=0$ is assumed for all $N_c\le t\le N_p$}. In the following we will reformulate ~\eqref{equ:MPC_ob_function} in a more straightforward manner. First, let \textcolor{black}{$\mathcal{Y}_t =\left[ y_{t+1}^{\top} \,\,y_{t+2}^{\top} \,\,\cdots \,\,y_{t+N_p}^{\top} \right] ^{\top}$}, one can compute
	\begin{equation}\label{equ:prediction_horizon}
	\mathcal{Y}_t =\varGamma \xi_t +\varTheta \varDelta U_t
	\end{equation}
	where
	\begin{equation}\nonumber
	\begin{aligned}
	\varGamma =&\left[ \mathscr{C}\bar{\mathscr{A}}\,\,\mathscr{C}\bar{\mathscr{A}}^2\,\,\cdots \,\,\mathscr{C}\bar{\mathscr{A}}^{N_p} \right] ^{\top} \\
	\varTheta =&\left[ \begin{matrix}
	\mathscr{C}\bar{\mathscr{B}}&		0&		\cdots&		0\\
	\mathscr{C}\bar{\mathscr{A}}\bar{\mathscr{B}}&		\mathscr{C}\bar{\mathscr{B}}&		\cdots&		0\\
	\mathscr{C}\bar{\mathscr{A}}^2\bar{\mathscr{B}}&		\mathscr{C}\bar{\mathscr{A}}\bar{\mathscr{B}}&		\cdots&		0\\
	\vdots&		\vdots&		\ddots&		\vdots\\
	\mathscr{C}\bar{\mathscr{A}}^{N_p-1}\bar{\mathscr{B}}&		\mathscr{C}\bar{\mathscr{A}}^{N_p-2}\bar{\mathscr{B}}&		\cdots&		\mathscr{C}\bar{\mathscr{A}}^{N_p-N_c}\bar{\mathscr{B}}\\
	\end{matrix} \right],
	\end{aligned}
	\end{equation}
	$G_t =\left[ g_{t}^{\top} \,\,g_{t+1}^{\top} \,\,\cdots \,\,g_{t+N_c-1}^{\top} \right] ^{\top}$, for $G=U$, $g=u$ and $G=\Delta U$, $g=\Delta u$.
	Therefore, we can reformulate optimization problem \eqref{equ:MPC_ob_function} as
	\begin{equation}\label{equ:mpc_optimization}
	\begin{aligned}
	\underset{\varDelta U_t ,\varepsilon}{\min}\ \ &J\left( \xi_t ,\varDelta U_t ,\varepsilon \right) =\left[ \Delta U_t ^{\top}\,\,\varepsilon \right] \mathcal{H}\left[ \Delta U_t ^{\top}\,\,\varepsilon \right]^{\top} +\mathcal{G}_t\left[ \Delta U_t ^{\top}\,\,\varepsilon \right]^{\top} \\
	&\ \ \ \ \ \ \ \ \ \ \ +\mathcal{P}_t  \\
	s.t.\ \ \ & \varDelta U_{\min}-\varepsilon \boldsymbol{1}_{mN_c}\le \varDelta U_t \le \varDelta
	U_{\max}+\varepsilon \boldsymbol{1}_{mN_c} \\
	& U_{\min}\le U_t \le U_{\max} \\
	& \varepsilon \ge 0
	\end{aligned}
	\end{equation}
	where $\mathcal{H}=\text{diag}\{\varTheta ^{\top}Q_{\mathcal{Y}}\varTheta +R_{\mathcal{Y}},		\rho\}$, $\mathcal{G}_t=\left[ \begin{matrix}
	2E_t ^{\top}Q_{\mathcal{Y}}\varTheta&		0\\
	\end{matrix} \right] $,
	$\mathcal{P}_t=E_t^{\top}Q_{\mathcal{Y}}E_t$
	and $E_t=\varGamma \xi_t - \mathcal{Y}_{ref,t}$, and $\mathcal{Y}_{ref,t}$ is defined as the sequence ${y}_{ref, t+1},{y}_{ref, t+2},\cdots ,{y}_{ref,t+N_p}$. The matrices $Q_{\mathcal{Y}}=\text{diag}\{Q,\cdots,Q\}\in\mathbb{R}^{NN_p\times NN_p}$ and $R_{\mathcal{Y}}=\text{diag}\{R,\cdots,R\}\in\mathbb{R}^{mN_c\times mN_c}$ .
	
	Let \textcolor{black}{$\Delta U_{t|t}=[\Delta u_{t|t}^{\top}\ \cdots\ \Delta u_{t+N_c-1|t}^{\top}]^{\top}$} be the optimal solution to~\eqref{equ:mpc_optimization} at time $t$, the control  applied to  the system  is
	
	\begin{equation}\label{equ:optimal_control}
	u_{t|t} =u_{t-1} +\varDelta u_{t|t}
	\end{equation}
	
	Then the optimization problem is solved again in the time $t+1$ according to the moving horizon strategy.
	To clearly illustrate the algorithm of DE-MPC, the implementing steps are summarized in  Algorithm \ref{alg:Deep_EDMD_MPC}.
	\begin{algorithm}[h]
		\caption{The DE-MPC Algorithm}
		\label{alg:Deep_EDMD_MPC}
		\begin{algorithmic}[1]
			\REQUIRE the trained $\theta_e$, $\theta_d$, $\mathscr{A}$, $\mathscr{B}$; prediction and control horizon $N_p$, $N_c$; weights matrices $Q$, $R$, $\rho$;
			\FOR {$i = 1, 2, \cdots$}
			\STATE Get the current lifted state $\xi_t:=\phi_e(x_t,\theta_e)$ and reference $\mathcal{Y}_{ref}$ using the planned trajectory according to (\ref{equ:encoder_parameterized});
%			\STATE Obtain the predicted lifted states $\mathcal{Y}$ according to (\ref{equ:prediction_horizon}) then acquire $E_t$ for calculating $\mathcal{G}_t$ and $\mathcal{P}_t$;
			\STATE Solve (\ref{equ:mpc_optimization}) to compute $\Delta U_{t|t}$;
			\STATE Compute $ u_{t|t}$ with~\eqref{equ:optimal_control} and apply it to the vehicle dynamical system.
			\ENDFOR
		\end{algorithmic}
	\end{algorithm}
	
	\section{Simulation and Performance Evaluation}
	In this section, the proposed Deep EDMD method was validated in a high-fidelity CarSim simulation environment. The comparisons with EDMD, ELM-EDMD, and DNN were considered to show the effectiveness of our approach.
	\begin{figure*}[h]
		\begin{center}
			\includegraphics[scale=0.9]{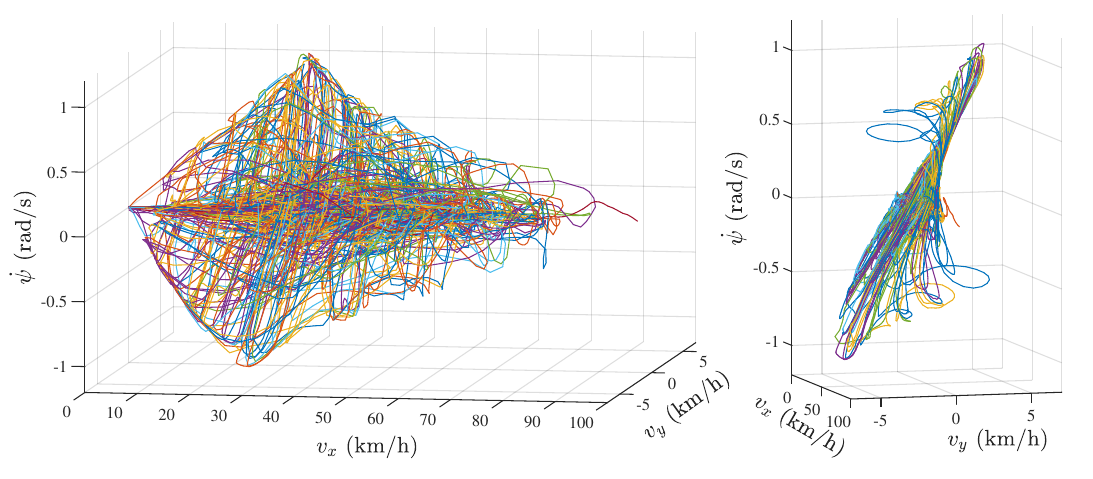}
			\caption{The distribution of the training data collected. Each path represents a complete episode collected.}
			\label{fig:training_data}
		\end{center}
	\end{figure*}
	
	\subsection{Data Collection}
	We validated our approach in a high-fidelity CarSim environment, where the dynamics were modeled with real vehicle data sets. Specifically, we chose a C-Class sedan car in the CarSim environment as the original Car model. The input-output data sets of the adopted vehicle were controlled with a Logitech G29 driving force steering wheels and pedals in CarSim 2019 %\textcolor{red}{on planar ground with friction coefficient 0.9}
	combined Simulink/MATLAB 2017b. The collected data sets consisted of 40 episodes and each episode filled with data of 10000 to 40000-time steps. In the data collection, the sampling period was chosen as $t_s=10 ms$. The initial state was set to zero for each episode, i.e., $x_0=0$. Besides, the engine throttle was initialized as $\eta_0=0$, while the steering wheel angle $\zeta_0$ was initialized randomly by the initial angle of the steering wheel entity. In the data collection process, we only allowed one of the throttle and the brake to control and the values of the throttle and brake are in the range of $[0\ 0.2]$ and $[0\ 9.1] \ MPa$. Besides, the value of the steering wheel angle was limited in the range of $[-450\degree\ 450\degree]$. Fig. \ref{fig:training_data} shows the distribution of the collected data sets, where each path represents a whole episode.
	\begin{table}
		\begin{center}
			\caption{Training hyperparameters in the simulations.}
			\scalebox{0.8}{
				\begin{tabular}{ccc}
					\toprule  %????????
					\textbf{Hyperparameter} & \textbf{Value}\\
					\midrule  %???????
					Learning rate & $10^{-4}$\\
					Batch size & 64 \\
					$L$ & 10 \\
					$p$ & 41 \\
					$\alpha_1$ & 1.0 \\
					$\alpha_2$ & 1.0 \\
					$\alpha_3$ & 0.3 \\
					$\alpha_4$ & $10^{-9}$ \\
					$\alpha_5$ & $10^{-9}$ \\
					$\alpha_6$ & $10^{-9}$ \\
					$\beta_1$ & 1.0 \\
					$\beta_2$ & $10^{-9}$ \\
					$\beta_3$ & $10^{-9}$ \\
					\bottomrule %????????
					\label{tab:training_hyperparameters}
			\end{tabular}}
		\end{center}
	\end{table}
	\subsection{Data preprocessing and training}
	The collected data sets were normalized according to the following rule
	\begin{equation}\label{equ:normalization}
	x =\frac{x-x_{min}}{x_{max}-x_{min}}
	\end{equation}
	
	We collected about $7.7 \cdot 10^5$ frame snapshots data {consisting} of 40 episodes and randomly chose 2 episodes as testing set, 2 episodes for validation, and the rest 36 episodes were regarded as the training set. In the training process, every two adjacent state samples were concatenated as new state data. Also, to enrich the diversity of the training data, we randomly generated the sampling starting point in the range of $[0 \ p]$ so that we could sample data sequences from a different starting position before each training epoch.
	\begin{figure*}[h!]
		% \vspace{-0.5cm}
		%	\setlength{\abovecaptionskip}{-0.1cm}
		%	\setlength{\belowcaptionskip}{-1cm}
		\begin{center}
			\includegraphics[scale=0.6]{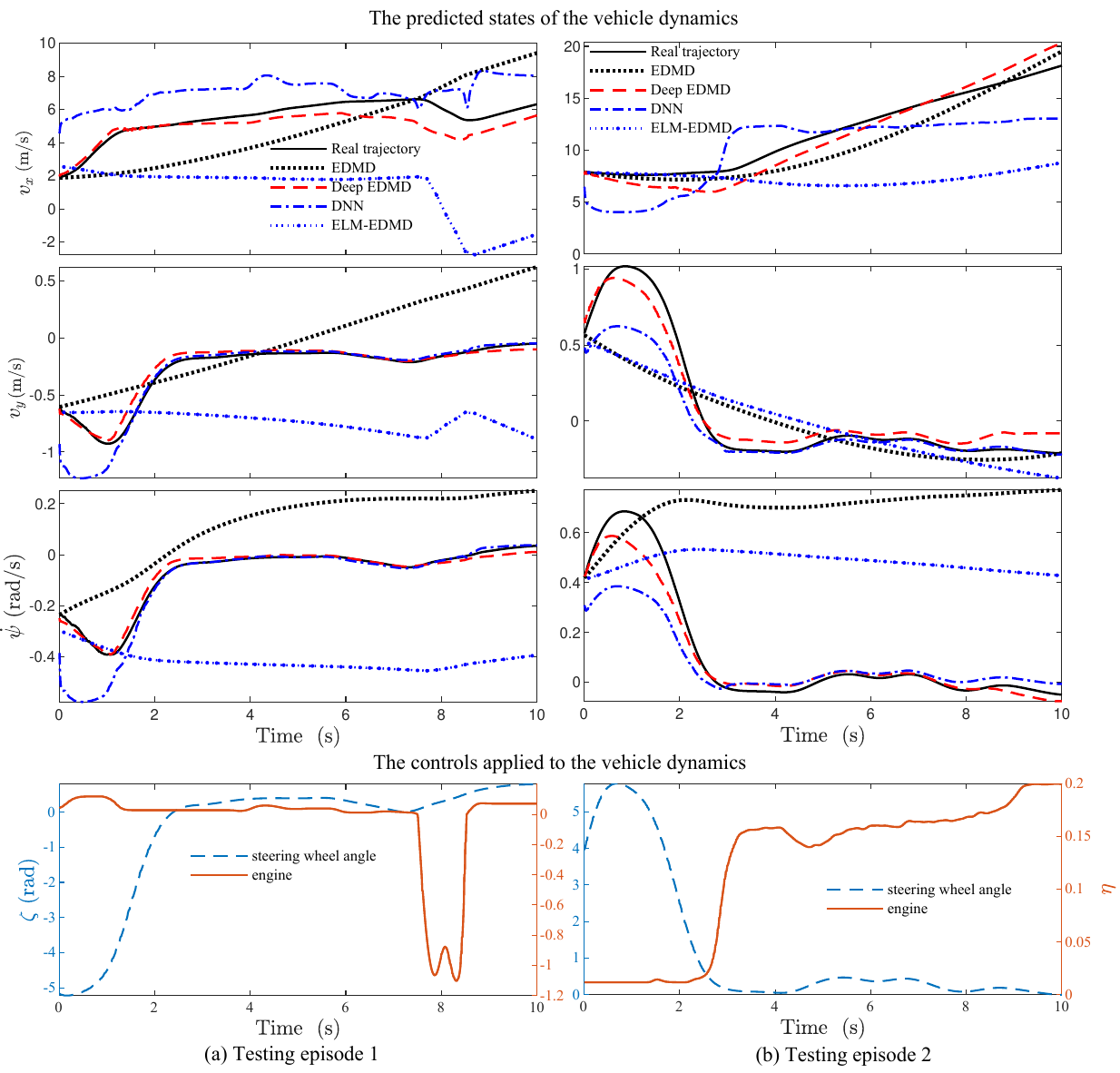}
			\caption{The predicted states of the vehicle dynamics and the applied controls. The upper three panels are the state predictions with Deep EDMD, EDMD, ELM-EDMD, and DNN, while the lowest panel shows the controls applied to the system.}
			\label{fig:normal_train_1}
		\end{center}
	\end{figure*}
	\begin{figure}[h!]
		% \vspace{-0.5cm}
		%	\setlength{\abovecaptionskip}{-0.1cm}
		%	\setlength{\belowcaptionskip}{-1cm}
		\begin{center}
			\includegraphics[scale=0.7]{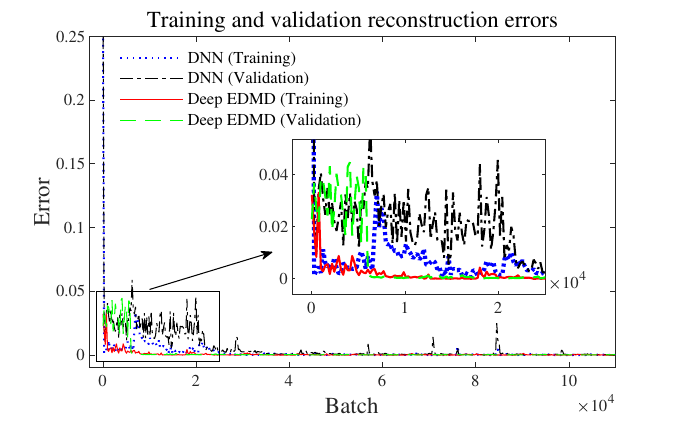}
			\caption{The training and validation reconstruction errors of Deep EDMD and DNN. The solid lines indicate the training errors, and the dashed lines indicate the validation errors.%The multi-step training and validation reconstruction errors in (\ref{equ:loss_recon}) of Deep EDMD and the training and validation loss in (\ref{equ:mlp_loss}) of the DNN.
			}
			\label{fig:h1}
		\end{center}
	\end{figure}
	In Deep EDMD, the structure of the encoder was of five layers with the structure chosen as $[n\ \ 32\ \ 64\ \ L\ \ L]$. And the structure of the decoder was set as $[L+n\ \ 128\ \ 64\ \ 32\ \ n]$. The simulations were performed with $L$ being set as $10$. All the parameters used in the simulation are listed in Table \ref{tab:training_hyperparameters}. All the weights were initialized with a uniform distribution limiting each weight in the range $[-f\ f]$ for $f=1/\sqrt{a}$, where $a$ is the number of the network layer \cite{goodfellow2016deep}.
	\begin{figure*}[h!]
		%	\vspace{-0.5cm}
		%	\setlength{\abovecaptionskip}{-0.1cm}
		%	\setlength{\belowcaptionskip}{-1cm}
		\begin{center}
			\includegraphics[scale=0.6]{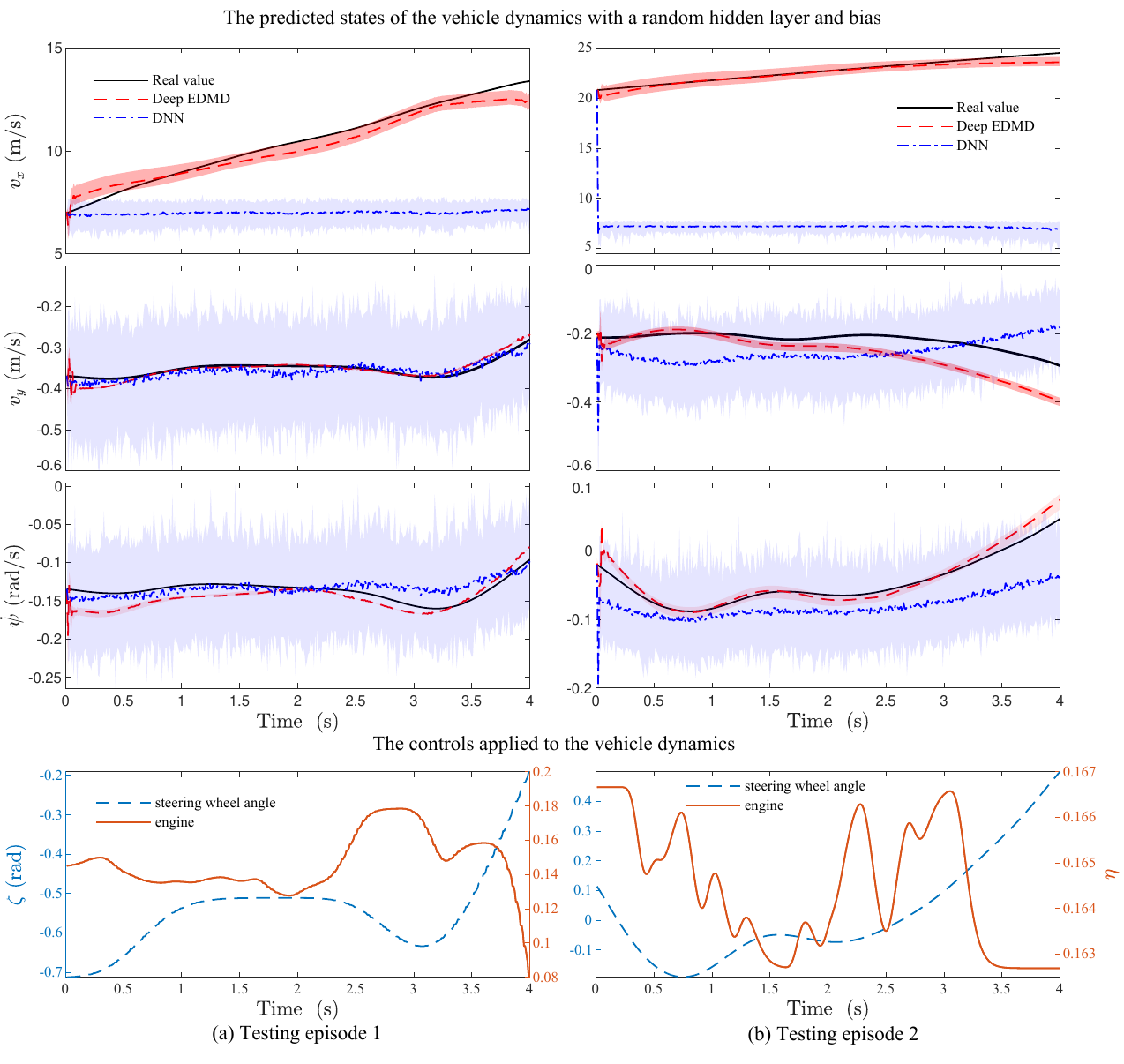}
			\caption{The resulting multi-step predictions with a random hidden layer and bias. (a) and (b) are testing episodes with different longitudinal velocities. The upper three panels shows the prediction results of the proposed Deep EDMD method and DNN with an additional random layer. The filled areas are the prediction errors and error variations of 100 repeated predictions. The lowest panel shows controls applied to the system.}
			\label{fig:abnormal_train_1}
		\end{center}
	\end{figure*}
	\begin{figure}[h!]
		% \vspace{-0.5cm}
		%	\setlength{\abovecaptionskip}{-0.1cm}
		%	\setlength{\belowcaptionskip}{-1cm}
		\begin{center}
			\includegraphics[scale=0.7]{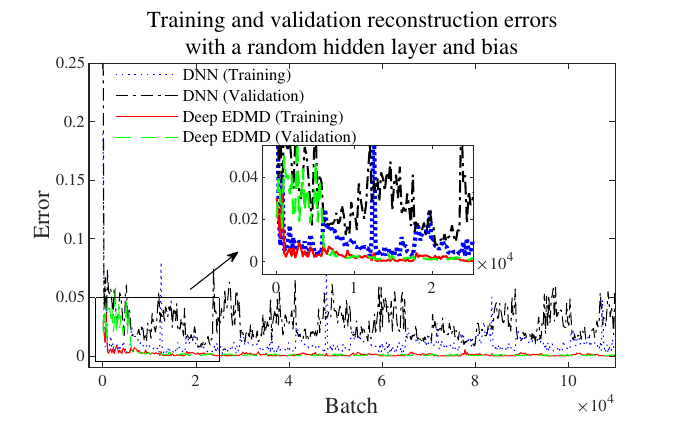}
			\caption{The training and validation reconstruction errors of Deep EDMD and DNN with an additional random hidden layer and bias. The solid lines indicate the training errors, and the dashed lines indicate the validation errors.}
			\label{fig:h2}
		\end{center}
	\end{figure}
	For comparison, an EDMD and an ELM-EDMD were designed to show the effectiveness of the proposed approach. In the EDMD method, we used the kernel functions described in \cite{cibulka2019data} to construct the observable functions. We adopted thin plate spline radial basis functions as the basis functions and the kernel centers were chosen randomly with a normal distribution $N\left(m, s^{2}\right)$
	\begin{equation}\label{equ:tpsrbf}
	\phi \left( x \right) =\lVert x-C \rVert ^2\log \left( \lVert x-C \rVert \right)
	\end{equation}
	where $C \in \mathbb{R}^{m \times (L+n)}$ are the centers of the basis functions $\phi$, $s$, and $m$ are the standard deviation and mean of the data sets. The ELM-EDMD method adopted the same structure as the EDMD approach, except for the kernel functions being replaced by the extreme learning machines for learning the observable functions.
	
	Also, a DNN method was used for comparison, to show the strength of our approach in terms of modeling and robustness.The structure of the DNN method was chosen as $[n+m\ \ 32\ \ 64\ \ 128\ \ 128\ \ 64\ \ 32 \  n]$. Two scenarios are considered in the comparison. In the first scenario, all the parameters were regarded as optimization variables to be learned, while in the second scenario, an additional hidden layer was added in the Deep EDMD method and the DNN method respectively, where weights and biases were updated randomly with a normal distribution during the whole training process. The loss function adopted in DNN was given as follows:
	\begin{equation}\label{equ:mlp_loss}
	\begin{aligned}
	L_{mlp}=&\frac{\beta _1}{p}\sum_{i=1}^p{\lVert x_{i+1}-M\left( \hat{x}_i \right) \rVert _{2}^{2}} +\frac{\beta _2}{p}\sum_{i=1}^p{\lVert x_i-\hat{x}_i \rVert _{\infty}}+\\
	&\beta _3\lVert \theta _M \rVert _{2}^{2}
	\end{aligned}
	\end{equation}
	where $\hat{x}_{i}=M(\hat{x}_{i-1})$ denotes the predicted next state, $\theta_M$ is the trainable weights and biases of DNN. The first term denotes the multi-step predicted loss, the second term is the infinite norm to penalize the largest error of the prediction, and the last term is the $l_2$ regularization term used for avoiding over-fitting. And the $\beta_1$, $\beta_2$ and $\beta_3$ are the weights of each term and their values are shown in Table \ref{tab:training_hyperparameters}.
	The EDMD method was trained in the MATLAB 2017b environment with an Intel i9-9900K@3.6GHz, while the methods of Deep EDMD, ELM-EDMD, and DNN were trained using the Python API in Tensorflow \cite{abadi2016tensorflow} framework with an Adam optimizer \cite{kingma2014adam}, using an NVIDIA GeForce GTX 2080 Ti GPU.
	\subsection{Performance evaluation}
	%To investigate the robustness of the training process, we compared our Deep EDMD with DNN about changes in the training errors in the training process. Besides, to demonstrate the identifying capability, we designed the contrast simulations with our Deep EDMD, DNN, EDMD, and ELM-EDMD on vehicle state prediction according to control sequences.
	
	\begin{comment}
	\begin{figure}[htbp]
	% \vspace{-0.5cm}
	\setlength{\abovecaptionskip}{-0.1cm}
	\setlength{\belowcaptionskip}{-1cm}
	\begin{center}
	\includegraphics[scale=0.7]{simulation_6.pdf}
	\caption{The predicted states of the vehicle dynamics and the applied controls: The upper three panels are the state predictions with the Deep EDMD, EDMD, ELM-EDMD, and the DNN, while the lowest panel shows the controls applied to the system.}
	\label{fig:normal_train_1}
	\end{center}
	\end{figure}
	\begin{figure}[htbp]
	\vspace{-0.5cm}
	\setlength{\abovecaptionskip}{-0.1cm}
	\setlength{\belowcaptionskip}{-1cm}
	\begin{center}
	\includegraphics[scale=0.7]{simulation_13.pdf}
	\caption{The predicted states of the vehicle dynamics and the applied controls. The upper three panels are the state predictions with the Deep EDMD, EDMD, ELM-EDMD, and the DNN, while the lowest panel shows the controls applied to the system.}
	\label{fig:nornal_train_2}
	\end{center}
	\end{figure}
	\end{comment}
	
	\textbf{Validation of the learned model with Deep EDMD:} For numerical comparison, the trained models with all the algorithms were validated with the same training data sets. The resulting state predictions of all the algorithms under the same control profile are illustrated in Fig. \ref{fig:normal_train_1}.  It can be seen that the resulting models Deep EDMD can capture the changes of the velocities in the longitudinal and lateral directions, which means the models are effective at a wide operating range. Also, the root mean square errors (RMSEs) of each algorithm corresponding to the trajectories in Fig. \ref{fig:normal_train_1} are listed in Table \ref{tab:RMSE_normal_cases}, which shows that the RMSEs of the proposed Deep EDMD for $v_x$, $v_y$, and $\dot{\psi} $ are smallest ones among all the approaches.
	In addition, in the two testing cases, the results obtained with Deep EDMD using $L=10$, are better than that with EDMD and ELM-EDMD. Specifically, the state trajectories obtained with EDMD and ELM-EDMD diverge after a short time period. It is worth noting that, DNN exhibits comparable performance with Deep EDMD in the prediction of lateral velocity and yaw rate, but shows larger prediction errors in the prediction of longitudinal velocity.
	
	To fully compare our method and DNN, the training and validation errors collected in the training process are shown in Fig. \ref{fig:h1}. It can be seen that, after 30000 batches of training, the training and validation reconstruction losses of Deep EDMD with $L=10$ converge to $2.4\cdot 10^{-4}$ and $3.8\cdot 10^{-4}$ respectively, and to $1.7\cdot 10^{-4}$ and $1.9\cdot 10 ^{-4}$ after 100000 batches. As for DNN, the training and validation losses converge to $8.9\cdot 10^{-4}$ and $4.7\cdot 10^{-3}$ after 30000 batches, and to $2.0\cdot 10^{-4}$ and $1.5\cdot 10^{-3}$ after 100000. The converging speeds of DNN are slower than that of our approach.
	
	\begin{comment}
	\begin{figure}[htbp]
	\vspace{-0.5cm}
	\setlength{\abovecaptionskip}{-0.1cm}
	\setlength{\belowcaptionskip}{-1cm}
	\begin{center}
	\includegraphics[scale=0.75]{simulation_random_4.pdf}
	\caption{The resulting multi-step predictions with a random hidden layer and bias. The upper three panels shows the prediction results of the proposed Deep EDMD and DNN with an additional random layer. The filled areas are the prediction errors and error variations of 100 repeated predictions. The lowest panel shows controls applied to the system.}
	\label{fig:abnormal_train_1}
	\end{center}
	\end{figure}
	\begin{figure}[htbp]
	\vspace{-0.5cm}
	\setlength{\abovecaptionskip}{-0.1cm}
	\setlength{\belowcaptionskip}{-1cm}
	\begin{center}
	\includegraphics[scale=0.75]{simulation_random_9.pdf}
	\caption{The resulting multi-step predictions with a random hidden layer and bias. The upper three panels shows the prediction results of the proposed Deep EDMD and DNN with an additional random layer. The filled areas are the prediction errors and error variations of 100 repeated predictions. The lowest panel shows controls applied to the system.}
	\label{fig:abnormal_train_2}
	\end{center}
	\end{figure}
	\end{comment}
	
	\begin{table*}
		\caption{{The RMSEs of the predicted states.}}
		\centering
		\begin{tabular}{ccccccccc}
			\toprule	
			%	\hline	
			& \multicolumn{4}{c}{Test episode 1} & \multicolumn{4}{c}{Test episode 2}\\
			\midrule
			%	\hline
			& EDMD & Deep EDMD & DNN & ELM-EDMD & EDMD & Deep EDMD & DNN & ELM-EDMD \\
			$v_x$ & 1.99 & \textbf{0.75} & 1.78 & 4.88 & 1.42 & \textbf{1.15} & 2.80 & 5.61\\
			$v_y$ & 0.36 & \textbf{0.04} & 0.14 & 0.55 & 0.25 & \textbf{0.08} & 0.16 & 0.27 \\
			$\dot{\psi}$ & 0.20 & \textbf{0.02} & 0.08 & 0.37 & 0.66 & \textbf{0.05} & 0.12 & 0.44 \\
			\bottomrule %????????
			%	\hline
			\label{tab:RMSE_normal_cases}
		\end{tabular}
	\end{table*}
	\begin{table}
		\caption{\textcolor{black}{The RMSEs of the predicted states in the case with random hidden layers in the Deep EDMD and DNN.}}
		\centering
		\begin{tabular}{ccccc}
			\toprule		
			& \multicolumn{2}{c}{Test episode 1} & \multicolumn{2}{c}{Test episode 2}\\
			\midrule
			& Deep EDMD & DNN & Deep EDMD & DNN \\
			$v_x$ & \textbf{0.4} & 3.86 & \textbf{0.35} & 15.56 \\
			$v_y$ & 0.01 & 0.01 &  \textbf{0.05} & 0.06\\
			$\dot{\psi}$ & 0.01 & 0.01 & \textbf{0.01} & 0.04 \\
			\bottomrule %????????
			\label{tab:RMSE_random_cases}
		\end{tabular}
	\end{table}
	\begin{table}
		\caption{\textcolor{black}{The RMSEs of  reference tracking with DE-MPC.}}
		\centering
		\begin{tabular}{ccc}
			\toprule		
			& $N_p=10, N_c=7$ & $N_p=60, N_c=50$\\
			\midrule
			$v_x$ & 1.38 & 0.93 \\
			$v_y$ & 0.20 & 0.18 \\
			$\dot{\psi}$ & 0.08 & 0.05 \\
			\bottomrule %????????
			\label{tab:RMSE_comtrol_cases}
		\end{tabular}
	\end{table}
	\begin{figure}[h]
		% \vspace{-0.5cm}
		%	\setlength{\abovecaptionskip}{-0.1cm}
		%	\setlength{\belowcaptionskip}{-1cm}
		\begin{center}
			\includegraphics[scale=0.8]{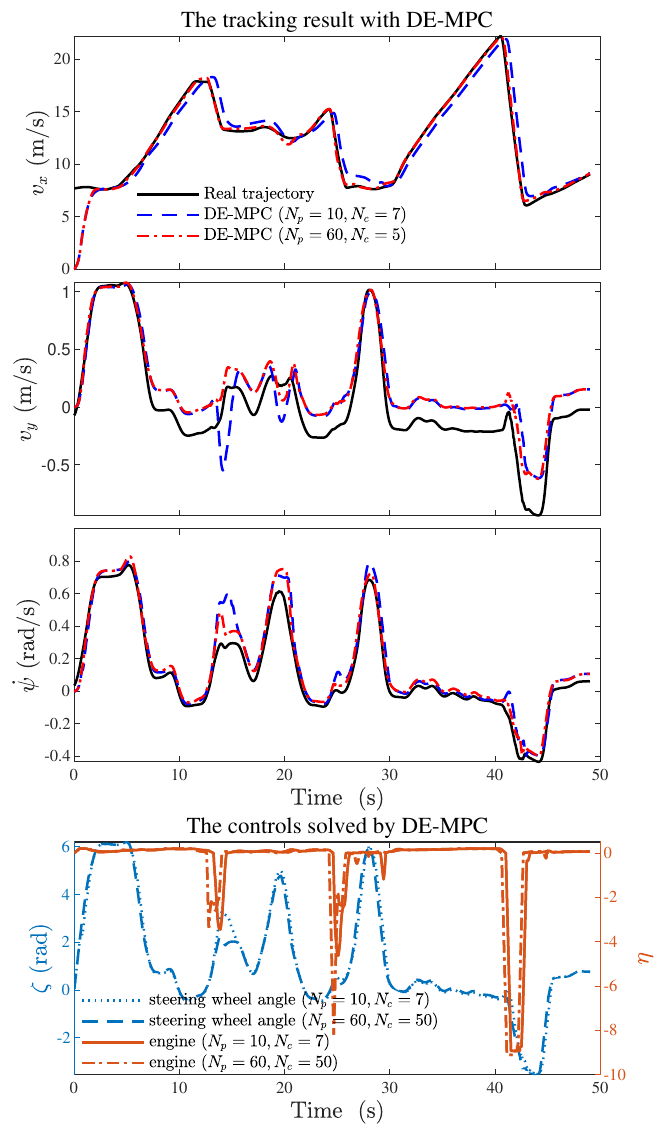}
			\caption{The tracking result with DE-MPC. We simulated with two different prediction horizon and control horizons, namely $N_p=10, N_c=7$ and $N_p=60, N_c=50$.}
			\label{fig:deep_EDMD_MPC_result}
		\end{center}
	\end{figure}
	In order to show the robustness of our approach, the obtained models of Deep EDMD and DNN with an additional random layer and bias, are used to generate the state predictions with 100 repeated times with different training data sets. The results of state predictions are collected and shown in Fig. \ref{fig:abnormal_train_1} \textcolor{black}{and the prediction RMSEs of Deep EDMD and DNN are collected in Table \ref{tab:RMSE_random_cases}}.
	The results show that the average prediction errors and error variations with Deep EDMD are much smaller than that with DNN. This is due to the Koopman operator with the EDMD framework adopted in the proposed approach. Indeed, the worst scenario of Deep EDMD with the random layer can be regarded as an EDMD framework, hence the robustness of the results can be guaranteed.
	
	We also collect the training and validation reconstruction errors obtained with Deep EDMD and DNN. The results are displayed in Fig. \ref{fig:h2}, which show that the random hidden layer has a large negative effect on DNN based training process. The associated training and validation errors fluctuate around 0.01 and 0.027. As for Deep EDMD with $L=10$, the training and validation reconstruction errors converge to $7.6\cdot 10^{-4}$ and $7.3\cdot 10^{-4}$ eventually, which are much smaller than that with DNN.
	
	\textbf{Trajectory tracking with DE-MPC}: To further validate the potential of the proposed Deep EDMD approach, we have designed the DE-MPC controller for tracking a time-varying reference in the CarSim/Simulink simulation environment.  In the simulation test,  the control steering wheel angle, throttle, and the brake pressure are limited in the interval  $[-450\degree\ 450\degree]$, $[0\ 0.2]$, and $[0\ 9.1] \ MPa$ respectively. Also, the control increments are limited respectively to the range $[-2.25\degree \ 2.25\degree]$, $[-0.004 \ 0.004]$, and $[-0.18\ 0.18]\ MPa$. The state of the vehicle has been initialized as $[0 \ 0\ 0]$ and $Q=1000I$, $R=5I$, $\rho = 10$. And the reference signals for the output have been selected from the collected testing data-set. The sampling interval in the simulation has been chosen as $t_s=10ms$.
	
	In order to fully show the effectiveness of the learned model, the simulation tests have been performed under two different prediction horizon  choices, i.e., $N_p = 10$, $N_c = 7$ and $N_p = 60$, $N_c = 50$. The simulation results for the control tests are shown in Fig. \ref{fig:deep_EDMD_MPC_result} \textcolor{black}{and the RMSEs of reference tracking under these two cases are listed in Table \ref{tab:RMSE_comtrol_cases}}. It can be seen that in both cases satisfactory tracking performance can be realized. In the case $N_p = 10$ and $N_c = 7$, a slight overshoot occurs when tracking the reference of $v_y$ and recovered in a number of steps, while the controller performs better in the case $N_p = 60$, $N_c = 50$  in terms of both tracking accuracy and smoothness. This is probably due to the greater prediction horizon and control horizon being used. Despite the larger prediction horizon adopted, the online computation is efficient since the linear model constraint is used. The average computational time for $N_p=10$ and $60$ are $0.0047s$ and $0.0092s$ respectively. As a consequence, the real-time implementing requirement can be satisfied considering the sampling interval being 10ms.
	
	\section{Conclusions}
	In this paper, we propose a novel data-driven vehicle modeling and control approach based on deep neural networks with an interpretable Koopman operator. In the proposed approach, a deep learning-based extended dynamic mode decomposition (Deep EDMD) algorithm is presented to learn a finite basis function set of the Koopman operator. Based on the dynamic model learned by Deep EDMD, a novel model predictive controller called DE-MPC is presented for path tracking control of autonomous vehicles. In the proposed algorithm, deep neural networks serve as the encoder and decoder in the framework of EDMD for learning the Koopman operator. The differences of our approach with the classic DNN for modeling vehicle dynamics can be summarized as follows: i) in our approach, the deep learning networks are designed as the encoder and decoder to learn the Koopman eigenfunctions and Koopman modes, which is different from classic machine learning-based modeling methods; ii) the resulting model is a global predictor with a linear dynamic evolution and a nonlinear static mapping function, hence it is friendly for real-time implementation with optimization-based methods, such as MPC.

	Simulation studies with data sets obtained from a high-fidelity CarSim environment have been performed, including the comparisons with EDMD, ELM-EDMD, and DNN. The results show the effectiveness of our approach and the advantageous point in terms of modeling accuracy than EDMD, ELM-EDMD, and DNN. Also, our approach is more robust {than} DNN in the scenario which has the random hidden and bias layer in the training process. We also tested the performance of DE-MPC for realizing trajectory tracking in the CarSim environment. Satisfactory tracking performance further verifies the effectiveness of the proposed Deep EDMD and DE-MPC methods. Future research will utilize the learned model for trajectory tracking in a real-world experimental platform of autonomous vehicles.
	\begin{comment}
	\section{CRediT authorship contribution statement}
	\textbf{Yongqian Xiao:} Methodology, Software, Writing- Original draft. \textbf{Xinglong Zhang:} Conceptualization, Methodology, Writing- Original draft. \textbf{Xin Xu:} Supervision, Writing- Original draft, Funding acquisition. \textbf{Xueqing Liu:} Software. \textbf{Jiahang Liu:} Investigation.
	\section{Acknowledgement}
	This work is supported by the National Natural Science Foundation of China under Grants 61751311, 61825305, the National Key R$\&$D Program of China	2018YFB1305105, and the Natural Science Foundation of Hunan Province of China under Grant 2019JJ50738.
	\end{comment}
	%\bibliographystyle{IEEEtran}
	%\bibliography{DeepEDMD}
	%*************************************************************************
%	\ifCLASSOPTIONcaptionsoff
%	\newpage
%	\fi
	
	%*************************************************************************
	%\end{CJK*}
	%*************************************************************************
	%******************************************************************
    \normalem
	\bibliographystyle{IEEEtran}
	\bibliography{IEEEabrv,DeepEDMD}

\end{document}